Jean-Philippe Rennard
jp@rennard.org

# Implementation of logical functions in the Game of Life

**Abstract**: The Game of Life cellular automaton is a classical example of a massively parallel collision-based computing device. The automaton exhibits mobile patterns, gliders, and generators of the mobile patterns, glider guns, in its evolution. We show how to construct basic logical operations, AND, OR, NOT in space-time configurations of the cellular automaton. Also decomposition of complicated Boolean functions is discussed. Advantages of our technique are demonstrated on an example of binary adder, realized via collision of glider streams.

The Game of Life (Life) is probably the most well known cellular automaton. The rules of its behavior were discovered by John Conway at the end of the 1960s. The presentation of Conway's construction by Martin Gardner in the October 1970 issue of Scientific American made it so famous that, in 1974, Time magazine even complained about how much computer time could be wasted because "growing hordes of fanatics" spent their office days playing with the new "toy". The newly discovered cellular automata rules were called "the Game of Life" because Conway " ... wanted to see some self-reproducing animal ... displaying some interesting behavior. In a weak form, living ... " [3]. He succeeded.

Obviously, the "living" is seen as a metaphor in the Game of Life context, also no spontaneous nontrivial self-reproducing patterns were found. However, the Game of Life possesses abilities to self-reproduction [6] as well as computational universality [5]. Actually, a simulation of the Game of Life universality was proved by its creator [1]. He showed that a universal Turing machine is embedded in the Game of Life, i.e. behavior of the Turing machine is imitated by space-time dynamic of the Game of Life cellular automaton. This chapter deals rather with computational, or logic, universality. We will present quite a simple way to implement any Boolean function in patterns of the Game of Life. All constructions discussed in the chapter are designed in LogiCell Java applet [4].

Sect. 1.1 offers a short survey of the Game of Life and introduces some basic patterns, that will be used in further constructions. In Sect. 1.2 we present logical gates and Boolean operators. Sect. 1.3 introduces particulars of collisions between mobile patterns in the Game of Life. Implementation of fundamental Boolean operators in space-time dynamics of the Game of Life is studied in Sect. 1.4. We show how to combine logical gates, in order to manage complex Boolean equations, in Sect. 1.5. The rest of the chapter deals with the solution of a simple combinatorial problem, the construction of a binary adder.

## 1.1 Basic Features of the Game of Life

The Game of Life is a two dimensional cellular automaton with binary cell states and Moore neighborhood. Each cell of the automaton takes either 0 or 1 state (we can call them living or dead, active or quiescent) and updates its state in discrete time depending on the states of its eight closest neighbors (Fig. 1.1a).

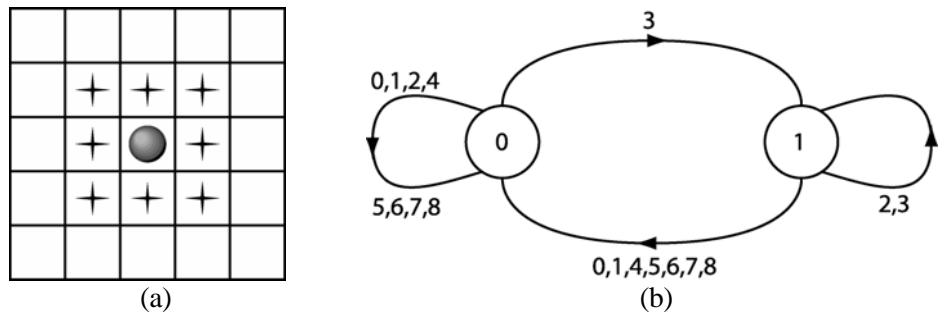

(a)                                    (b)

**Fig. 1.1**. The Game of Life basics. (a) Cell neighborhood: central cell is a disc and its eight closest neighbors are "+"s. (b) Cell state transition: nodes of the graph are states of a cell, arcs symbolize cell state transitions, they are labeled by numbers of neighbors that cause the transition.

Even though the Game of Life can easily be redefined in a totalistic manner, it has been originally designed as a semi-totalistic automaton. When a cell updates its state, it does not look at the exact configuration of its neighborhood but only considers the number of active neighbors. Also, the cell itself is excluded from its neighborhood. Thus, cells of the Game of Life obey the three simple rules:

Rl:    A dead (0) cell with exactly three active neighbors becomes alive (1).
R2:    A living (1) cell with two or three active neighbors remains alive (1).
R3:    In any other case the cell dies or remains quiescent/dead.

Essentially, each cell is a finite state automaton, therefore the cell state transition rules can be expressed in a classical automaton state transition diagram (Fig 1.1b).

Using terms of population dynamic one can say that for a cell to be born its local density of population must be large enough (the rule Rl). The cells die of loneliness (less than two neighbors) or overpopulation (more than three neighbors), the rule R3. The rules generate quite a complex, not "mechanical", dynamic of the cellular automaton, where patterns of "alive" cells grow and change their appearance in an unpredictable way. S. Wolfram considered the Game of Life as a typical example of a complex cellular automaton, belonging to the class IV of its hierarchy of cellular automata [7].

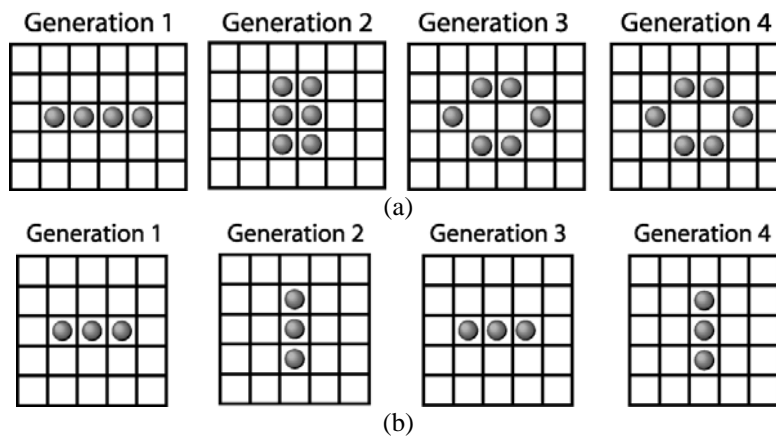

**Fig. 1.2.** Two simple immobile patterns in the Game of Life. (a) Formation of a stable pattern (beehive). (b) An oscillating pattern (blinker).

When we fill a lattice with 0s and 1s at random and allow the configurations to develop for a while we find that after some time most patterns inhabiting the lattice can be classified as follows:

- Stable patterns. They remain unchanged if not disturbed. The simplest one is the 2 x 2 cell square block. Another example is the generation of a beehive (Fig. 1.2a).

- Blinkers. They usually change their configuration in a very short cycle. For example, in Fig 1.2b, we can see a three-cells pattern, blinker, oscillating between horizontal and vertical forms.

In studies of the Game of Life a particular pattern, R-pentomino (Fig. 1.3a), was discovered. It was thought to be a minimal initial configuration that generates unpredictably developing patterns; the R-pentomino stabilizes only after 1103 generations. Actually, the R-pentomino generates a cluster of small stable or oscillating patterns, and also produces six mobile patterns (Fig. 1.3b), that run away from the R-pentomino original position as photons from the Sun. This mobile pattern consists of five non-zero states, the configuration of which is changed in cycle to translate the pattern along the lattice (Fig. 1.3c). The glider runs with the speed of $c/4$, where $c$ is what Conway called the "speed of light", one cell jump.

Gliders belong to a class of spaceships, compact mobile patterns, that can be used as signals, or information quanta, in collision-based computing devices. Actually, spaceships were considered essential for the design of a universal Turing machine and there should be a pattern that produces spaceships. The last one, known as a glider gun, was discovered at the end of 1970 by R. Gosper. Its p30 glider gun is shown in Fig. 1.4. This pattern generates a glider every 30th step of discrete time. During more than a quarter of a century hundreds of glider guns, with various periods of gliders emissions, have been found. Traditionally, we will deal with the p30 gun as a core pattern in our constructions of logical gates.

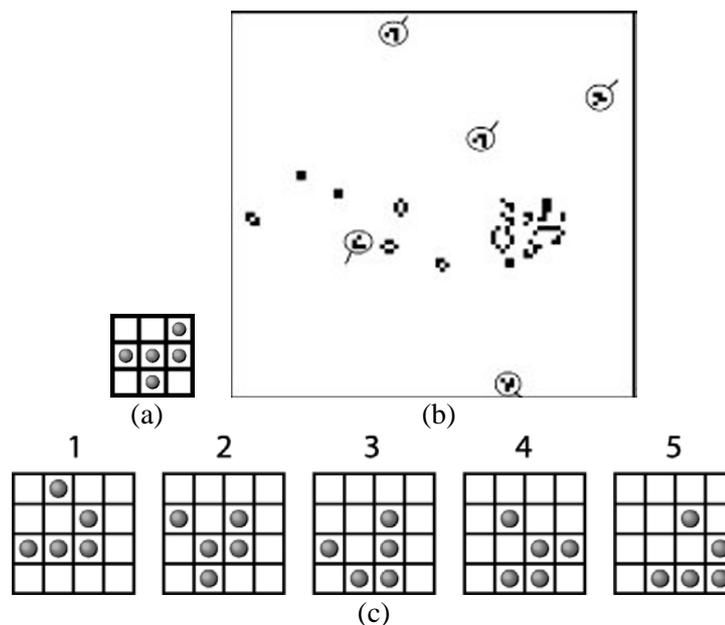

**Fig. 1.3.** The glider and its historical precursor. (a) Pentomino. (b) Configuration developed from the pentomino, at the 224th step of evolution; the gliders are encircled, and their velocity vectors are sketched. (c) The glider.

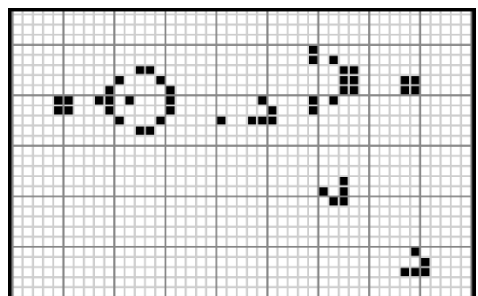

**Fig. 1.4**. The p30 glider gun.

## 1.2 Logical Gates

A logical gate is some kind of "black box" which is able to process two Boolean variables, inputs, according to a specified Boolean operator. There are three main gates, which correspond to the three fundamental operators defined by George Boole.

The AND operator is true if both inputs are true. The notation is *a* ^ *b*.

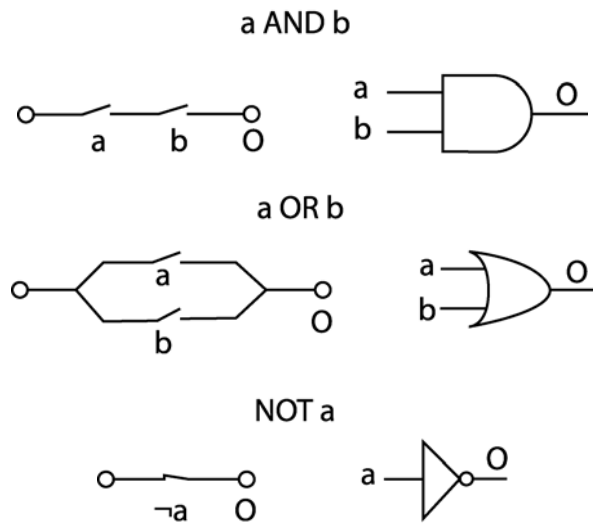

**Fig. 1.5.** Diagrams of the AND, OR and NOT gates.

The electrical diagram of the AND-gate (Fig. 1.5, left) shows that both switches *a* and *b* must be ON to activate the output O. The classical electronic diagram is presented too (Fig. 1.5, right): a gate associates two electrical inputs and processes the corresponding output.

The OR operator is true if at least one input is true. The notation is: *a* v *b*. The electrical diagram of the OR operator (Fig. 1.5, left) shows that to transmit the current to the output, only one switch has to be ON.

The NOT operator is unary, it uses only one entry and reverses it. The notation is ¬*a*. In the electrical diagram one uses an inverted switch which transmits the current when it is OFF. If one switches the entry ON, the output will no longer be activated.

Every logical function, i.e. every possible result set of the combination of two Boolean variables, can be constructed using these three fundamental operators. We then only have to implement AND, OR, NOT-gates to be able to manage any Boolean function. To implement a logical gate we therefore need:

- Some kind of electrical pulses to represent inputs.
- Wires to transmit the electrical pulses.
- Processing devices which associate inputs and compute the Boolean result.
- A device placed after the processing device, able to check the output electrical pulses. This will represent the output.

We thus encode these items in the Game of Life objects as follows:

- Input and output electrical pulses >> Gliders.
- Wires >> Trajectories of glider movements.
- Processing devices >> Collision of gliders.
- Output device >> Collision of gliders with immobile patterns.

## 1.3 Collision Reactions

Particulars of glider collisions underpin designs of all logical devices, constructed in the Game of Life automaton.

### 1.3.1   Glider Collisions

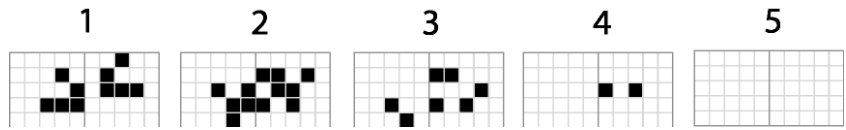

**Fig. 1.6.** An example of glider collision.

   Gliders annihilate when they collide with each other. In Fig. 1.6 we find an example of how gliders disappear when they are involved in 90° binary collision.
   If we place a right gun, i.e. a gun that emits gliders rightwards, and a left gun side by side, their streams of gliders will cancel each other as a result of glider collisions. This happens if the two following conditions are satisfied. The distance between nascent gliders is even (this is 32 in Fig. 1.7). There is a one-cell vertical offset between the glider streams (Fig. 1.7).
   When gliders are at the same level vertically when they collide with each other, the process of glider streams annihilation is two-phase. Firstly, collision of two gliders generates two 2 x 2 blocks (Fig. 1.8, 1-6). The gliders following the two previous ones crash into the blocks (Fig. 1.8, 23-27). The blocks and gliders disappear as a result of the collision.

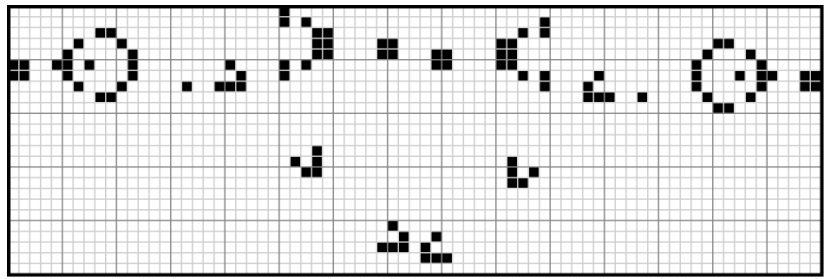

**Fig. 1.7.** Head on disposition of glider guns.

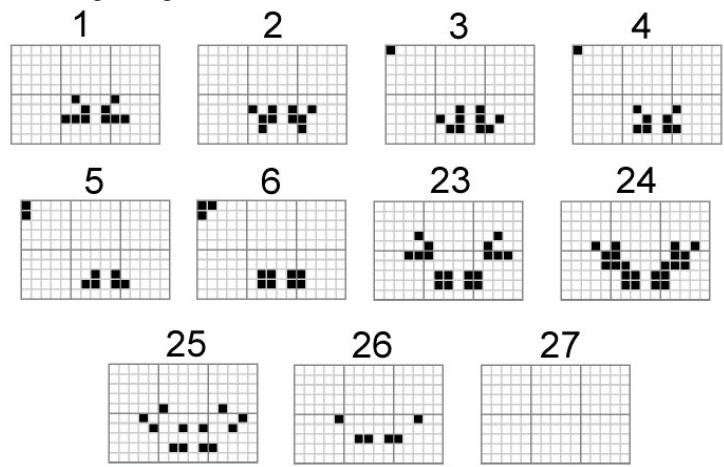

**Fig. 1.8.** Two step process of gliders' disappearance.

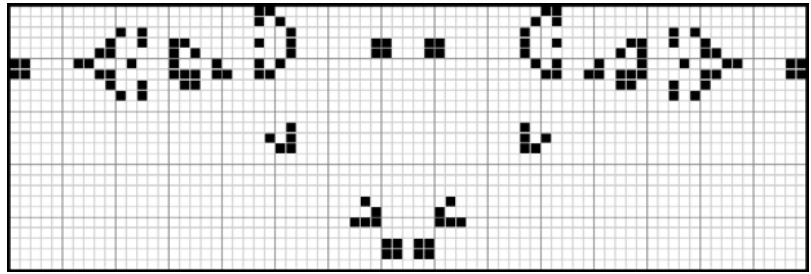

**Fig. 1.9.** The position of guns that eases control of glider streams.

There are some perturbations which arise during glider collisions (e.g. see Fig. 1.8, 25). To prevent oncoming gliders from being bit by "debris" of the previous collision we need to keep quite a big distance between the gliders in the stream. This is the case with the p30 gun (Fig. 1.9). Consequently an adequate positioning of guns allows for the efficient control of streams of gliders.

### 1.3.2   Eaters

To design efficient circuits we need to somehow stop a stream of gliders to prevent the gliders from "polluting" the computational space. There are compact stable patterns, called eaters that consume gliders and then recovery back to their original form.

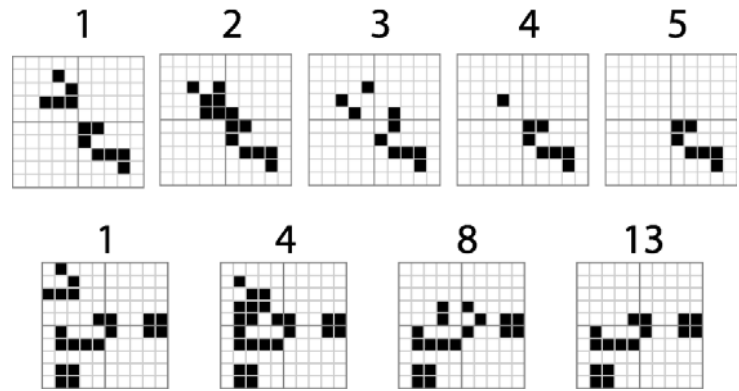

**Fig. 1.10.** Two types of eaters that consume the glider. Upper snapshots (1-5) show collision of a glider with a stopper. Bottom snapshots (1-13) demonstrate detection of a glider by another eater.

In Fig. 1.10 we can find the two forms of eaters utilized in our further constructions. The first one will be used as a stopper to prevent glider streams from further expansion. The second eater will be employed to detect presence of gliders at a given site of the space.

### 1.4 Implementation of Logical Gates

In the previous sections we acquired basics of logic gate designs and got some clues to utilize glider collisions. Let us now discuss implementation of basic components - inputs, outputs, and gates - in configurations dynamics of the Game of Life automaton.

### 1.4.1 Input

First of all we must represent electrical pulses going along input wires. The pulses can be implemented with a glider gun that generates a stream of gliders. Obviously, the wires themselves will not be "physically" build, they are simply the gliders trajectories. Generally speaking, a stream of gliders can encode any data. For example, the number "101" will be a series "glider - no glider - glider". We also want to control glider streams according to the data they carry. Like in an electrical circuit, the gliders must propagate only if the input is true. We must therefore find a way to stop the gliders if the input is false. This can be done via coupling the glider gun with a stopper. As we have seen in the previous sections the eaters eliminate gliders and therefore terminate the glider streams. So, to feed input data into our computing device we destroy stoppers situated in specific positions. A stopper can be easily destroyed by activation, i.e. switching to the state 1, of a resting cell, closest to it. An example, of such "external" activation of the system is shown in Fig. 1.11. The eater disappears in nine steps (Fig. 1.12) when this specific cell is activated.

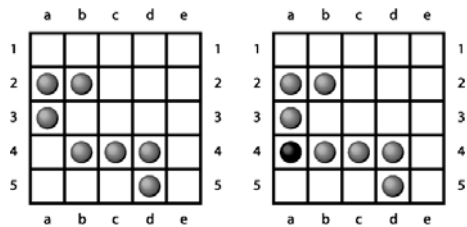

**Fig. 1.11.** The input activation. The entry cell with coordinates a-4 is colored dark.

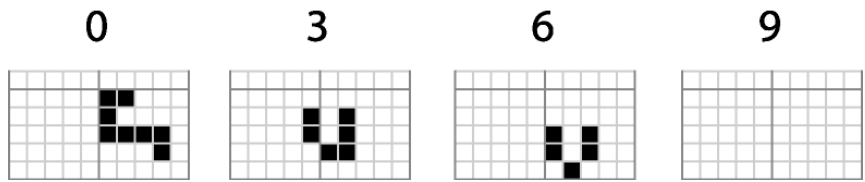

**Fig. 1.12.** Disappearance of the activated stopper.

Positioning the eater carefully relative to the gun, we can stop the gliders unless the input is activated. When the input is activated, the stopper vanishes opening the way for the signal (Fig. 1.13). Each input therefore could be represented by such an association of specific patterns, constituting an element of the logical gates (component). If there are more than one instance of a given input in an equation, we only have to build one component per instance of the input. The initial values of course will be identical for each of them.

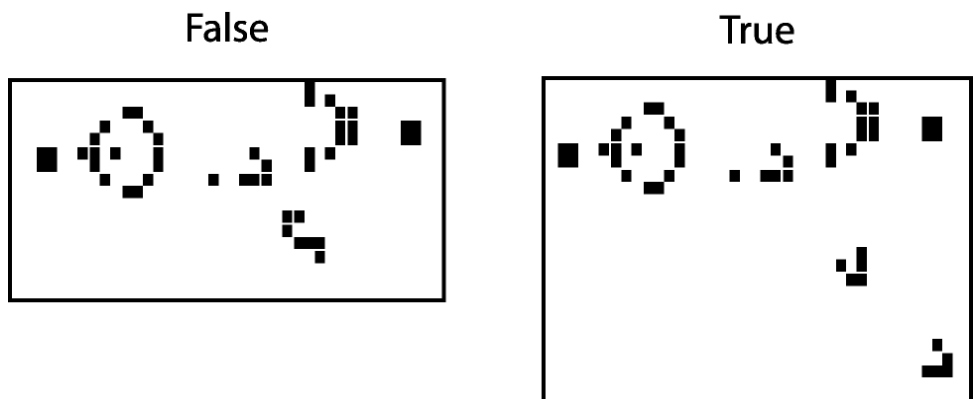

**Fig. 1.13.** An entry component after 60 steps.

### 1.4.2   Output

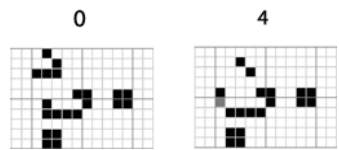

**Fig. 1.14**. Output activation. The light cell is one specifically activated.

If we place an eater on the trajectory of a glider some specific cells are activated only when a collision occurs (Fig. 1.14). It is not difficult to compute the periodicity of the gate, i.e. the number of discrete time steps necessary for the signal to join the output. Therefore, to obtain the output value we simply have to check a state of one specific cell, as e.g. the light cell in the Fig. 1.14. If the cell is activated the output value is considered to be true, otherwise it is false. We could have used a stopper to represent the output, however we employed an eater of a different type to make overall construction transparent.

## 1.5 Coupling the Components

In order to build a logical gate we will need the following four components, which are based on four different patterns: input, simple p30 guns, stoppers, and outputs. By combining inputs and guns it is possible to control the streams of gliders generated by the inputs. Stoppers are used only to prevent the streams of gliders from propagating outside the gate. The output is used to check the presence of gliders that result from the computation. Assuming the Game of Life cellular automaton is used as a hardware device, the only links between the program or interface and the cellular automaton are the following:

- The positions of the input cells. There is one input cell per input in the logical gate.
- The position of the output cell. There is of course only one output per gate.
- The periodicity of the gate.

To solve a Boolean equation, the interface sets the input cells according to the Boolean variables values and checks then the output cell at the corresponding generation.

Now we are ready to build basic Boolean operators via coupling inputs with glider collisions and representing results as glider streams.

### 1.5.1   The AND-Gate

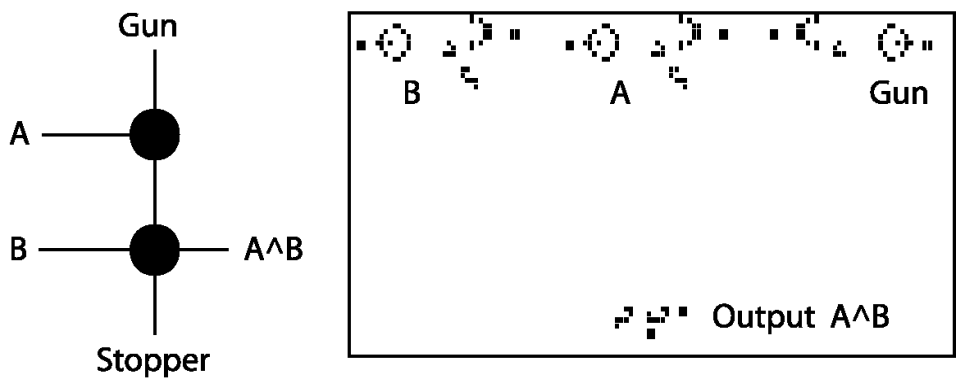

**Fig. 1.15.** Implementation of the AND-gate in the Game of Life. There is a gate's scheme on the left and space-time dynamic, that realizes the gate, on the right.

The AND-gate requires two input components, *A* and *B*, an output *O* and a simple gun *G*. The structure of the gate is shown in Fig. 1.15.

The *B* gliders will move forward and join the output only if both *A* and *B* input variables are true. The *A* and *B* components are aligned side by side, the gun is perpendicular to the inputs, and the output is positioned on the trajectory of *B* gliders. A stopper is placed opposite the gun to stop the gun's gliders if both inputs are false. Thus, the Game of Life implementation of the AND-gate exhibits the following input-driven configurations (Fig. 1.16):

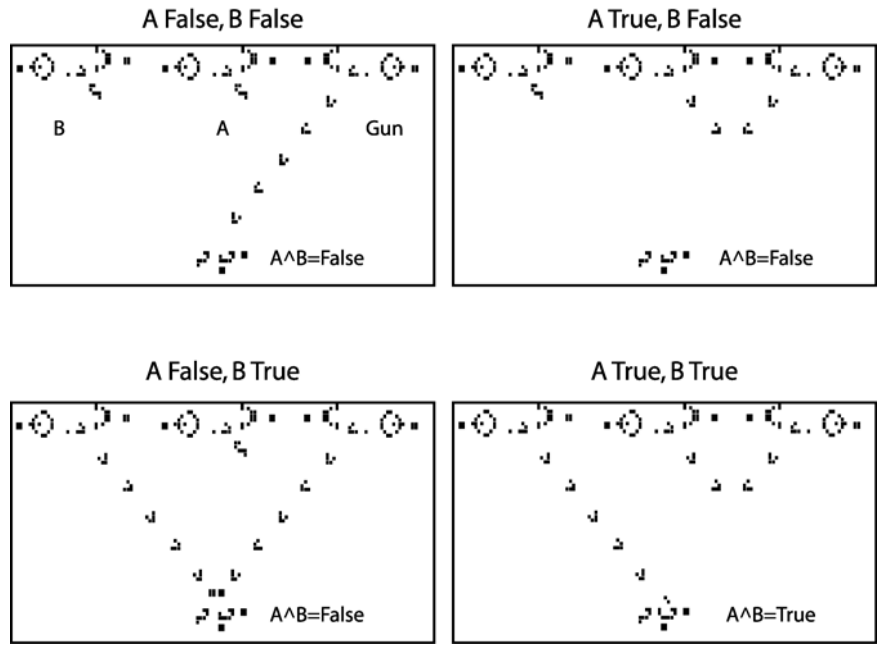

**Fig. 1.16**. Configurations of automaton for various input values of the AND-gate.

- *A* and *B* are both false: no glider can activate the output, the result is false.
- *A* is true and *B* is false: *A* stops the gun. Since *B* is false it cannot activate the output.
- *A* is false and *B* is true: *B* cannot join the output since the gun stops it.
- *A* and *B* are true: *A* stops the gun and *B* can activate the output. The result is true.

### 1.5.2  The OR-Gate

The OR-gate is slightly more complex. Obviously, like the AND-gate, it requires two input components and an output. However, now we need two guns, one parallel to the input and one perpendicular. The general structure of the gate is shown in Fig. 1.17.

The gliders emitted by the parallel gun will join the output unless both inputs are false. The second gun, perpendicular to the three others structures is placed near *A*. A stopper is positioned on the trajectory *B* to stop the *B* gliders when *A* is false. The output resides on the trajectory of the gliders emitted by the parallel gun. The configurations of the gun implementation are following (Fig. 1.18):

- *A* and *B* are both false: the perpendicular gun stops the parallel gun from activating the output. The result is false.
- *A* is true and *B* is false: *A* stops the gliders leaving the perpendicular gun, the parallel gun is free to activate the output. The result is true.
- *A* is false and *B* is true: *B* stops the perpendicular gun, therefore the parallel gun is free to activate the output. The result is true.
- *A* and *B* are true: *A* stops the gliders emitted by the perpendicular gun, the stopper stops *B,* therefore the parallel gun can activate the output. The result is true.

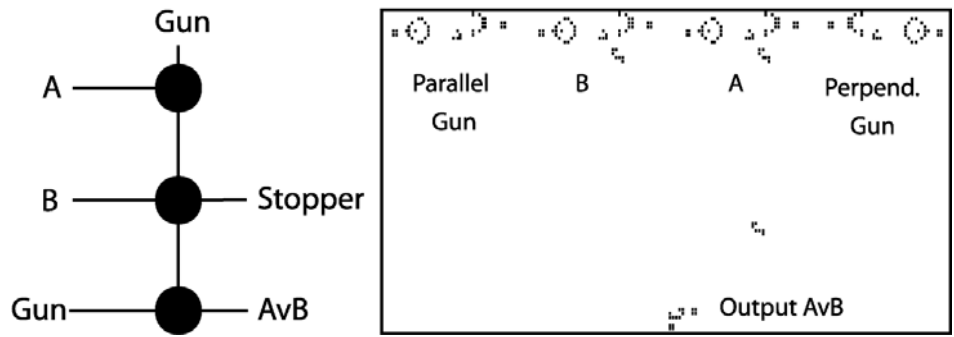

**Fig. 1.17**. Implementation of the OR-gate in the Game of Life. There is a gate's scheme on the left and space-time dynamic, that realizes the gate, on the right.

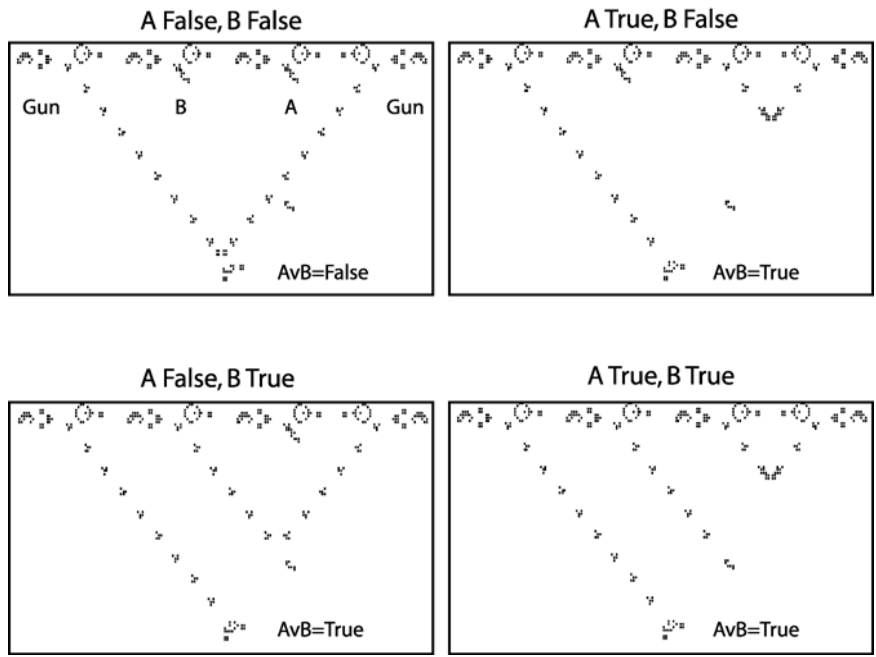

**Fig. 1.18**. Configurations of the automaton for various input values of the OR-gate.

### 1.5.3   The NOT-Gate

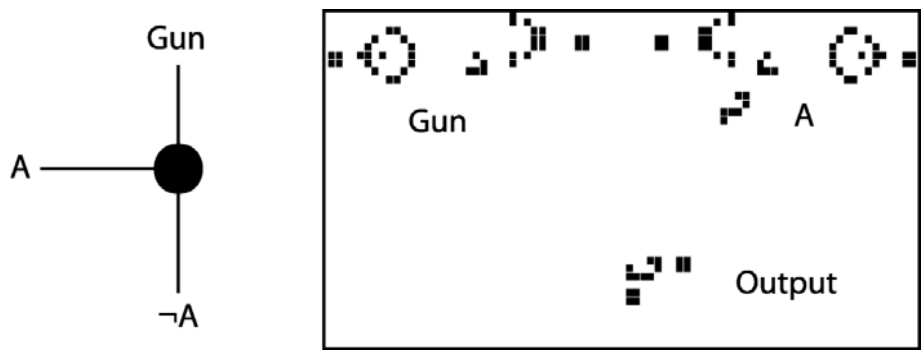

**Fig. 1.19**. Implementation of the NOT-gate in the Game of Life. There is a gate's scheme on the left and space-time dynamic, that realizes the gate, on the right.

Since NOT is an unary operator, its implementation is the simplest one. We only need an input component, a perpendicular gun and an output (Fig. 1.19). The gun will activate the output unless *A* prevents it from doing so.

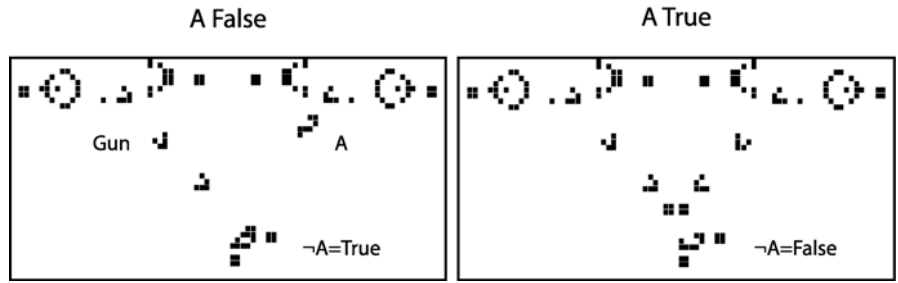

**Fig. 1.20**. Configurations of the automaton for various input values of the NOT-gate.

The configurations of the automaton implementation are as follows (Fig. 1.20):

- *A* is false: The gun is free to activate the output. The result is true.
- *A* is true: *A* does not allow the gun to activate the output. The result is false.

The NOT-gate does not simply reverse the input, it also reverses the direction of the activating glider, i.e. of the output. The construction of a non-inverting NOT-gate is possible but more complicated. We will see later how to manage the inversion when combining Boolean operators.

1.6 Implementation of Boolean Equations

To implement Boolean equations in the Game of Life automaton we need to couple the gates and to manage the NOT operation properly.

### 1.6.1   Gates Associations

Our gates only manage two inputs. To accept more inputs in a given equation we have to break the equation down in such a way that we always have to consider only two inputs at once. That is we decompose the original equation onto several simple sub-equations interconnected via their results. The technique is widely known and is particularly familiar in conveyor computing and automata models of compilers. However, we discuss it anyway to make the chapter self-contained.

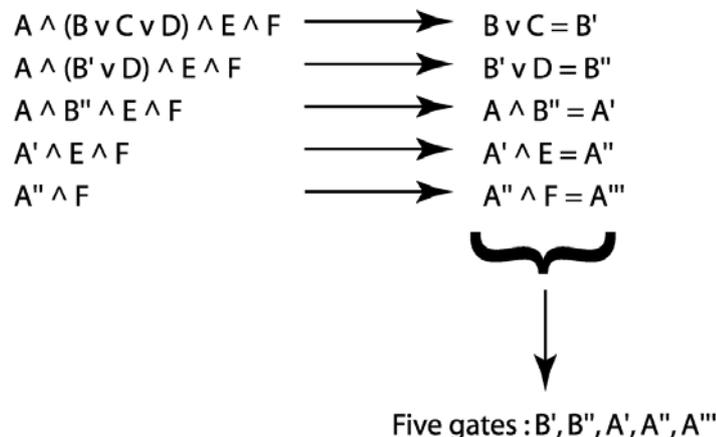

**Fig. 1.21**. Association of gates.

The AND and OR operators are associative, therefore it is always possible to pair the Boolean variables in the equations. Consider, for example, the simple equation $A \wedge B \wedge C$. It can be interpreted as $(A \wedge B) \wedge C$. Let $(A \wedge B) = A'$, we have to implement $A' = A \wedge B$ and then $A' \wedge C$. The recursive application of this method helps to reduce the number of input problems at a time step (Fig. 1.21). Therefore to implement any equation we only have to combine gates. Outputs of the high level gates are used as inputs of the lower level gates, e.g. $B'$ is one of the inputs of $B' \wedge D$. This means that for all the gates which level is greater than 0, we must remove the output eater. Then, according to the sizes and positions of these gates, we just have to compute the positions of the components of the lower level.

To build $A \wedge B \wedge C$ we should firstly design the gate $A \wedge B$ (without its output eater) and then combine its output, which is on the trajectory of $B$, with C (Fig. 1.22). The gate $A' = A \wedge B$ is colored gray in the Fig. 1.22.

As soon as we know the trajectories of the gate's output gliders, it is not difficult to position correctly the components of the gate $A' \wedge B$. In that case the output of the gate $A \wedge B \wedge C$ is on the $C$ entry trajectory. If $A' = A \wedge B$ is true the gun of $A' = A \wedge B \wedge C$ will be prevented from activating the output (Fig. 1.23).

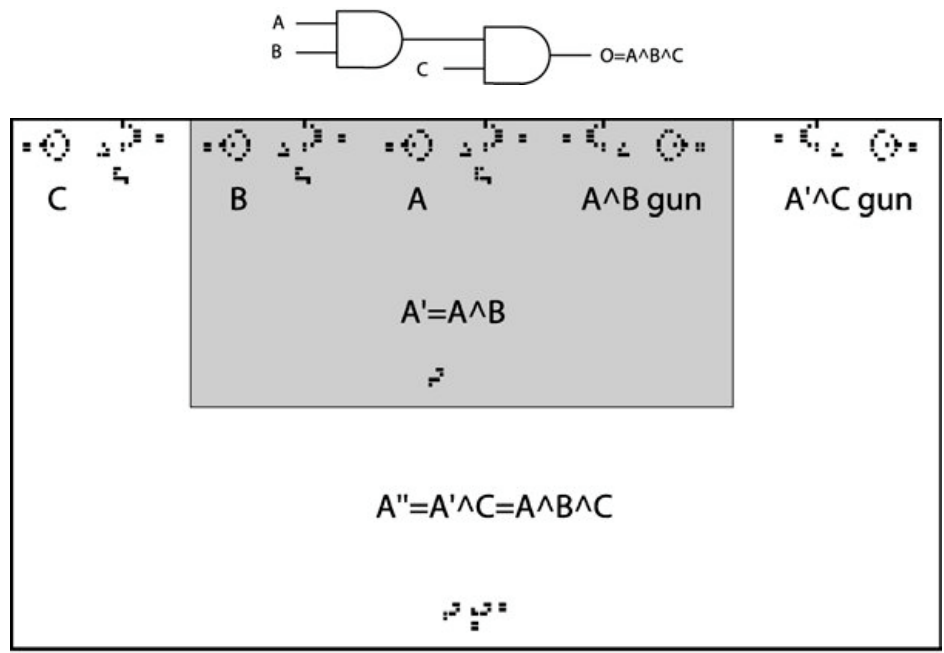

**Fig. 1.22**. Implementation of the function $A \wedge B \wedge C$ (up) in space-time dynamic of the Game of Life automaton (bottom).

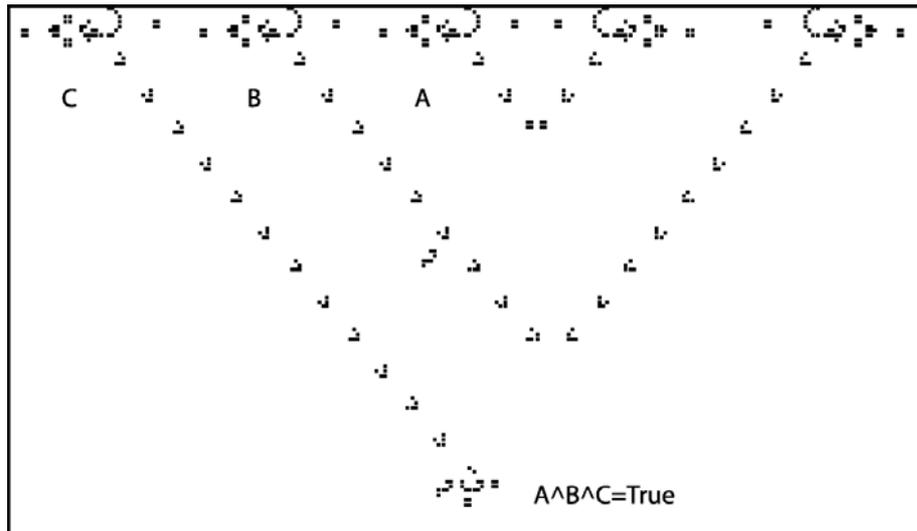

**Fig. 1.23**. Configuration of the automaton that implements the function $A \wedge B \wedge C$ all arguments of which take the value true.

### 1.6.2  Management of the NOT-Gate

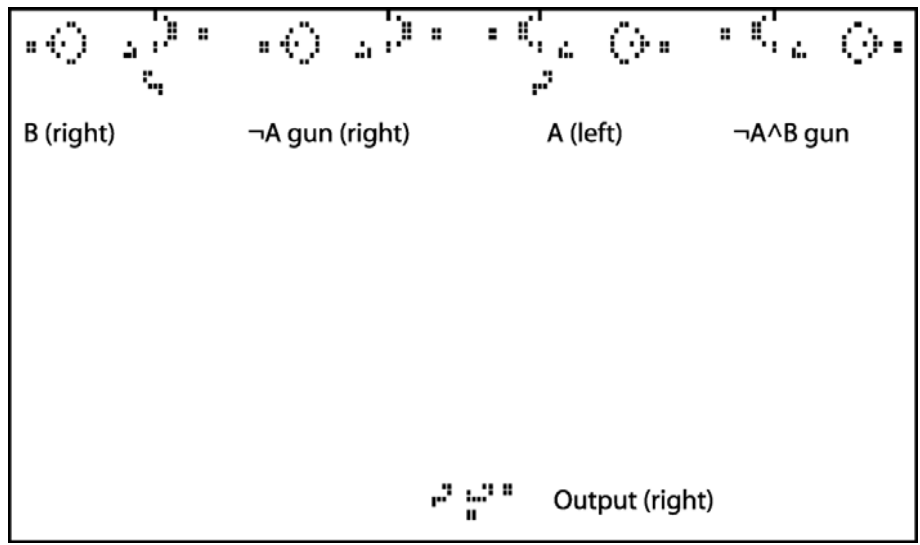

**Fig. 1.24**. Configuration of the automaton that implements the function $\neg A \wedge B$.

Most components of our constructions have a right side output: the gliders, that activate the output, move rightward. The NOT-gate we have presented reverses the output: a right side input results in a left side output. Since we have to combine all three operators, we must manage the direction problem. Considering that the direction of any component can be reversed, thanks to the totalistic rules of the Game of Life, this is easy to realize. Given a desired output direction, we only need to recursively reverse inputs according to the NOT operators. Let us consider the equation $\neg A \wedge B$. The output is on the *B* trajectory. Since we want a right side output, *B* has to be right sided. It must be combined with the $\neg A$ output which therefore must be right sided too. To obtain this, we only have to reverse the side of the *A* entry, if *A* is left sided, $\neg A$ becomes right sided (Fig. 1.24).

On this basis we are able to manage any Boolean equation, the only limitations being the computer memory and the response time. Furthermore we can implement many equations on the same lattice. Those equations will all be solved in parallel. To illustrate these capabilities let us examine the example of the following section.

## 1.7 Binary Adder

Now we are familiar with the implementation of logical gates and are therefore eager to solve some combinatorial problems. A binary adder is quite a good example. Let us consider a sum of one-bit numbers $x$ and $y$. The results of the addition can be written as follows:

```
    x   y     Sum  Carry
    -----------------------
    0   0      0    0
    1   0      1    0
    0   1      1    0
    1   1      0    1
```

The one bit sum $s$ represents the value of $x$ XOR $y$ ($x \omega y$). The XOR operator is true if one and only one of both variables is true. The carry bit $c$ represents $x \wedge y$.

Let $x$ and $y$ be $n$-bit values represented as follows:

$x_n x_{n-1} \ldots x_0$ and $y_n y_{n-1} \ldots y_0$

Then

$s_0 = x_0 + y_0$ and $s_i = x_i + y_i + c_{i-1}$

where $c_i$ is the carry bit of the sum $x_i + y_i$, and $i = 1 \ldots n$. We have

$s_0 = x_0 \omega y_0$

$s_i = x_i \omega y_i \omega c_{i-1}$

and

$c_i = (x_i \wedge y_i) \vee ((x_i \omega y_i) \wedge c_{i-1})$,

where $i = 1 \ldots n$.

The result is expressed on $n + 1$ bits ($b_{n+1} b_n \ldots b_1 b_0$) where $b_i = s_i$ for $i = 0 \ldots n$ and $b_{n+1}$ is the carry bit $c_n$. For example, if $x$ and $y$ are two-bit values then we have the following:

```
       x₁   x₀
  +    y₁   y₀
       ----------------
  b₂   b₁   b₀
```

The three bits of the result are

$b_0 = x_0 \omega y_0$

$b_1 = y_1 \omega x_1 \omega (x_0 \wedge y_0)$

and

$b_2 = c_1 = (x_1 \wedge y_1) \vee [(x_1 \omega y_1) \wedge (x_0 \wedge y_0)]$

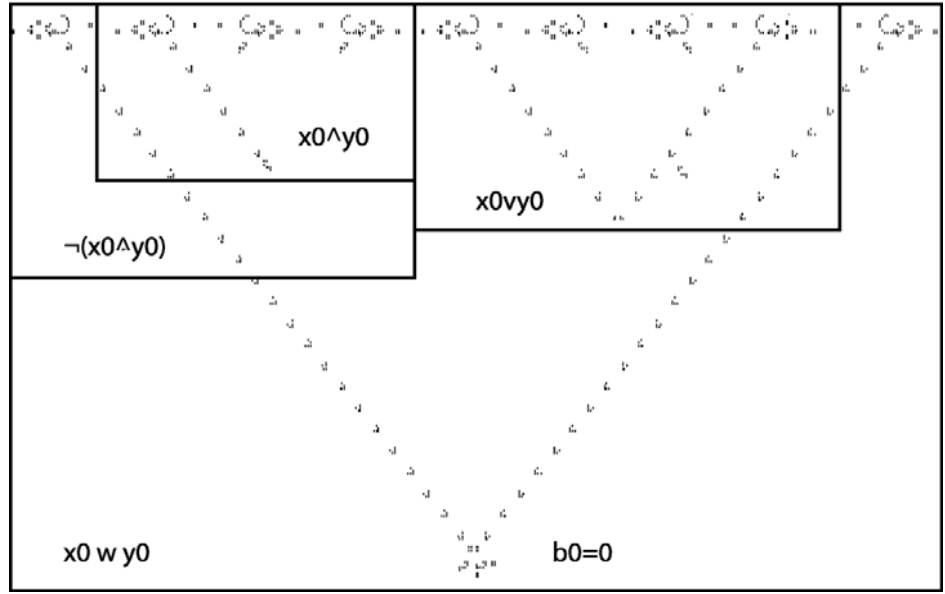

**Fig. 1.25.** Configuration of the automaton computing $b_0$ for the addition $10 + 10$.

Since we did not directly implement the XOR operator we must construct it with the three fundamental operators. XOR is classically defined as

$$x \, \omega \, y = (x \wedge \neg y) \vee (\neg x \wedge y)$$

but we need ten guns to represent this equation. If we use the representation

$$x \, \omega \, y = (x \vee y) \wedge \neg(x \wedge y)$$

then we only need nine guns. We therefore have (Fig. 1.25):

$$b_0 = x_0 \, \omega \, y_0 = (x_0 \vee y_0) \wedge \neg(x_0 \wedge y_0)$$

One could notice that since we want to combine $\neg(x_0 \wedge y_0)$ with $(x_0 \vee y_0)$ we must reverse the direction of $(x_0 \wedge y_0)$. It is more difficult to deal with the second bit (Fig. 1.26)

$$b_1 = y_1 \, \omega \, x_1 \, \omega \, (x_0 \wedge y_0) = (((y_1 \vee x_1) \wedge \neg(y_1 \wedge x_1)) \vee (x_0 \wedge y_0)) \wedge$$
$$\neg(((y_1 \vee x_1) \wedge \neg(y_1 \wedge x_1)) \wedge (x_0 \wedge y_0)).$$

The last bit is as follows (Fig. 1.27):

$$b_2 = (x_1 \wedge y_1) \vee [(x_1 \, \omega \, y_1) \wedge (x_0 \wedge y_0)] =$$
$$(x_1 \wedge y_1) \vee [((x_1 \vee y_1) \wedge \neg(x_1 \wedge y_1)) \wedge (x_0 \wedge y_0)].$$

We then only have to map all three equations onto the same cellular automata lattice to obtain a proper implementation of the binary adder. The equations are computed in parallel (Fig. 1.28). The result of $11 + 11$ is $110$.

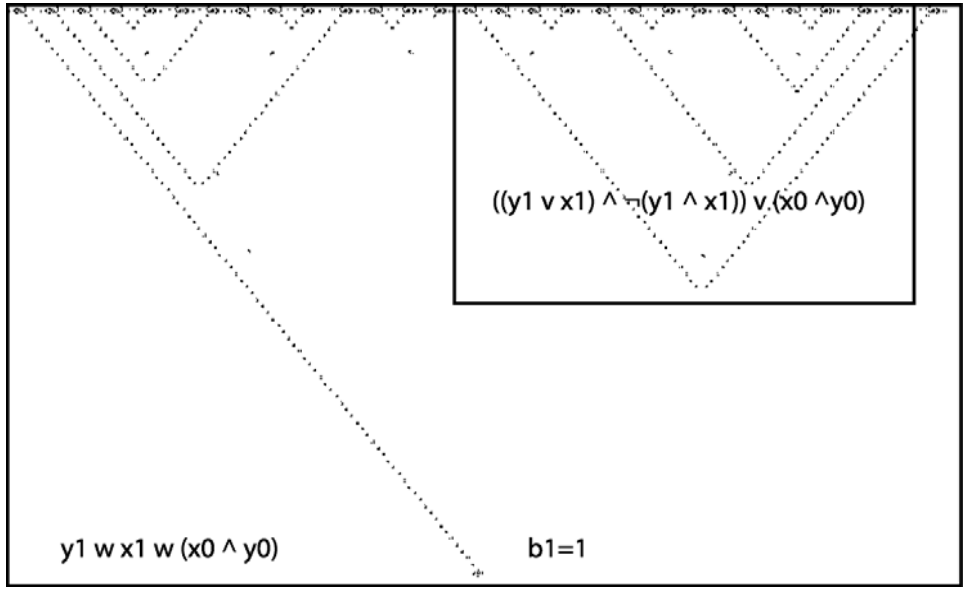

**Fig. 1.26**. Configuration of the automaton computing $b_1$ for $10 + 01$.

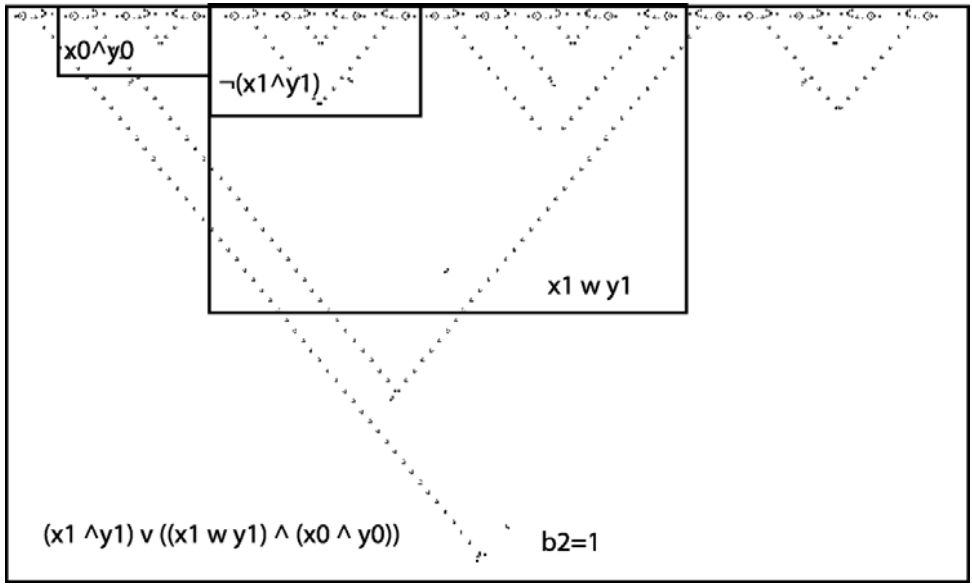

**Fig. 1.27**. Configuration of the automaton computing $b_2$ for $11 + 11$.

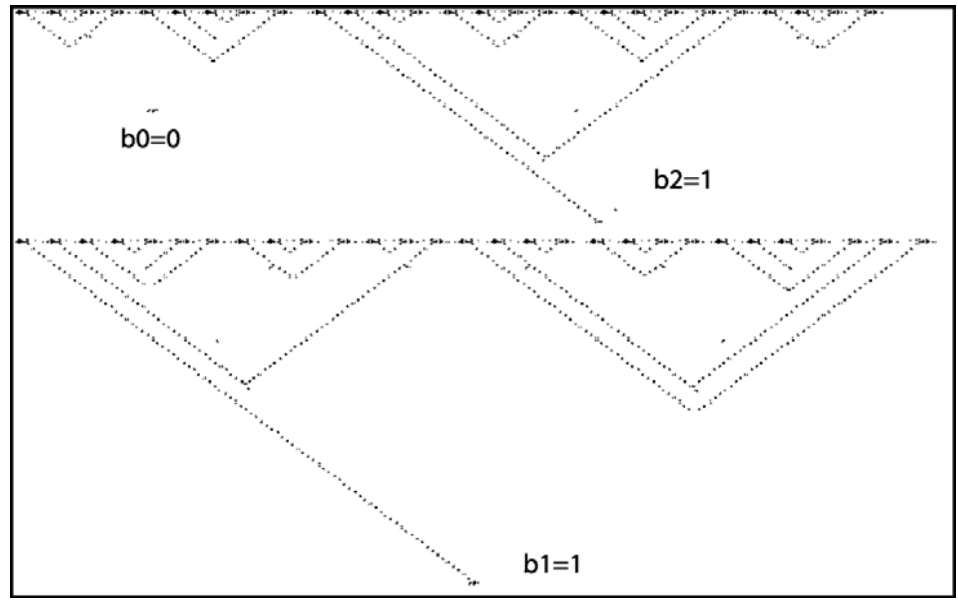

**Fig. 1.28**. Configuration of the automaton implementing binary adder for 11 + 11.

1.8 Conclusion

In this chapter we have shown how basic logical gates and functions can be implemented via collision of gliders in the Game of Life cellular automaton. We tried to keep ideas, underlying our constructions, as simple and intuitive as possible. As a result the built objects are understandable but rather non-efficient. Since the distance between the gliders of entry components is greater than the guns periods, the function of glider-based circuit response time $r$ can be seen as a function $r = f(2d, n + 2a + 3o, b)$ of the distance $d$ between the gliders of the guns and of input components, the numbers $n$, $a$, $o$ of NOT, AND and OR operators in problem equation, and the value $b$ characterizing offsets, or positions of the output components. When logical computation in the Game of Life cellular automaton is simulated in Java, in LogiCell software [4], 400 MHz Pentium II, the response time for the two bits binary adders is circa 2.5 minutes. Actually, the model can be greatly sped up. Some useful constructions, including copies and complementing the signals without turning glider streams and implementation of a delay operation, can be found in [2].

References

1. Berlekamp E.R., Conway J.H. & Guy R. *Winning Ways for Your Mathematical Plays, vol* 2 (Academic Press, 1982). The structures of our basic gates are introduced in this text.
2. Durand B. & Roka Z. The Game of Life: universality revisited *Research Report 98-01* (Ecole Normale Supéérieure de Lyon, Laboratoire de l'Informatique du Paralléélisme, 1998).
3. Levy S. *Artificial Life. The quest for a new creation* (Penguin Books, 1992).
4. LogiCell Java applet. http://www.rennard.org/alife
5. Morita K. & Imai K. A simple self-reproducing cellular automaton with shape encoding mechanism In: Artificial Life V (MIT Press, Cambridge, 1997) 489-496. 6. von Neumann J. (Burks A., Editor) *Theory of Self-Reproducing Automata* (University of Illinois Press, 1966).
7. Wolfram S. Universality and complexity in cellular automata *Physica D **10*** (1984) 1-35.